\documentclass[apj,11pt]{emulateapj}
\usepackage{xspace}
\usepackage{amsmath}
\usepackage{epsfig}
\usepackage{graphicx}
\usepackage{hyperref}

\def\nW{$\mathrm{nWm^{-2}sr^{-1}}$}
\def\mic{$\mu {\rm m}$}
\def\NLF{$342$}
\def\gsim{\gtrsim}
\def\lsim{\lesssim}

\begin{document}

\pagestyle{myheadings} \markright{DRAFT: \today\hfill}

\title{ Reconstructing the $\gamma$-ray Photon Optical Depth of the Universe to \lowercase{$z \sim 4$} from Multiwavelength Galaxy Survey Data }

\author{Kari Helgason$^{1,2}$ and Alexander Kashlinsky$^{2,3}$ }
\affil{$^1$Department of Astronomy, University of Maryland, College Park, MD 20742, USA; kari@astro.umd.edu}
\affil{$^2$Observational Cosmology Laboratory, Code 665, NASA Goddard Space Flight Center, Greenbelt MD 20771}
\affil{$^3$ SSAI, Lanham, MD 20706; alexander.kashlinsky@nasa.gov}
\email{kari@astro.umd.edu}

\begin{abstract}

We reconstruct $\gamma$-ray opacity of the universe out to $z\lsim 3-4$ using an extensive library of \NLF\ observed galaxy luminosity function (LF) surveys extending to high redshifts. We cover the whole range from UV to mid-IR (0.15-25\mic) providing for the first time a robust empirical calculation of the $\gamma\gamma$ optical depth out to several TeV. Here, we use the same database as \citet{Helgason12} where the extragalactic background light was reconstructed from LFs out to 4.5\mic\ and was shown to recover observed galaxy counts to high accuracy. We extend our earlier library of LFs to 25\mic\ such that it covers the energy range of pair production with $\gamma$-rays (1) in the entire {\it Fermi}/LAT energy range, and (2) at higher TeV energies probed by ground-based Cherenkov telescopes. In the absence of significant contributions to the cosmic diffuse background from unknown populations, such as the putative Population III era sources, the universe appears to be largely transparent to $\gamma$-rays at all {\it Fermi}/LAT energies out to $z\sim 2$ whereas becoming opaque to TeV photons already at $z\lsim 0.2$ and reaching $\tau \sim 10$ at $z=1$. Comparison of the currently available {\it Fermi}/LAT gamma-ray burst and blazar data shows that there is room for significant emissions originating in the first stars era.

\end{abstract}


\section{Introduction}

The extragalactic background light (EBL) supplies opacity for propagating high energy GeV-TeV photons via an electron-positron pair production ($\gamma\gamma\rightarrow e^+e^-$) \citep{Nikishov62,Stecker71}. Determining the transparency of the universe is of fundamental importance for a wide variety of current observatories such as the space-borne {\it Fermi}/LAT instrument operating at energies $\lsim 250-300$ GeV to ground-based $\gamma$-ray telescopes probing energies $\gsim$1 TeV. The distance at which the optical depth due to this interaction is $\tau\sim 1$ defines a horizon of the observable universe at $\gamma$-ray energies, and has been a subject of extensive efforts designed to model the build-up of EBL with time from the posited emission history of galaxy populations \citep[e.g.,][]{Stecker06,Franceschini08,Kneiske&Dole10,Dominguez11}.

In this Letter we show that, with observed galaxy populations over a wide range of wavelengths, one can uniquely reconstruct the optical depth of the universe at these energies out to redshifts $z \sim 4$. This empirical reconstruction relies exclusively on {\it data} from an extensive library of galaxy luminosity functions (LFs) encompassing 18 finely sampled wavelengths from UV to mid-IR (0.15-24 \mic) relevant for the pair-production opacity. This methodology enables robust calculation of the $\gamma$-ray opacity in the {\it Fermi}/LAT energy range using galaxy surveys probing $\lambda \leq 4.5 $\mic\ out to $z\lsim 4$. Extending to TeV energies, probed by the ground-based Cherenkov observations, we use measurements out to 24 \mic; this extrapolation is robust for the redshifts currently probed these observations.  This heuristic reconstruction using the {\it observed} galaxy populations defines the absolute floor of the photon-photon optical depth due to known galaxy populations and deviations from it would allow the characterization of any emissions inaccessible to direct telescopic studies \citep{Kashlinsky05a,Gilmore12b}.

We use the methodology developed in \citet{Helgason12} of reconstructing the EBL from observed galaxy populations in a compilation of \NLF\ measured LFs covering the UV, optical and near-IR; that compilation is slightly updated compared to Table 1 of \citet{Helgason12}. The wealth of galaxy survey data has recently reached adequate redshift coverage to make such empirical estimation of the evolving EBL feasible and the reconstruction was shown to reproduce independent data from galaxy counts and the cosmic infrared background \citep[CIB, ][]{Kashreview}. Our approach is completely independent of theoretical modeling describing the evolution of galaxy populations in that we use the LF data directly observed at all wavelengths out to $z\sim3-8$ in this heuristic reconstruction from which we derive the optical depth due to pair production \citep[see also][]{Stecker12}.

Standard cosmological parameters are used below: $\Omega_{\rm tot}=1$, $\Omega_{\rm m}=0.3$, $H_0=70 {\rm km \cdot s^{-1} Mpc^{-1}}$.

\section{Reconstructing the Evolving EBL from Data}

\begin{figure}
  \begin{center}[t]
      \includegraphics[width=0.45\textwidth]{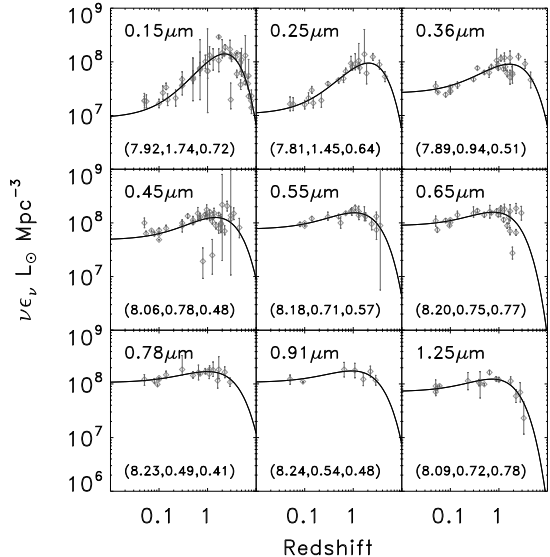}
  \end{center}
      \caption{ Measurements of the evolving luminosity density in our binned rest-frame wavelength range 0.14-1.25\mic\ in units of $L_\odot {\rm Mpc^{-3}}$. The solid curves are fits through the data points according to our fitting function (Equation~\ref{eqn:ldfit}) with best-fit parameters displayed in each panel as ($\log_{10} a_\lambda$,$b_\lambda$,$c_\lambda$) where $a_\lambda$ is also in $L_\odot{\rm Mpc^{-3}}$. All data points have been converted to units with $h=0.7$. The references for the data points can be found in Table 1 of \citet{Helgason12} with the addition of UV data from \citet{Steidel99,Sullivan00,Treyer05,Schiminovich05,Budavari05,Sawicki06,Finkelstein12}. }
      \label{fig_ld1}
\end{figure}

Quantifying the optical depth of the universe to high energy photons requires knowledge of the properties of the intervening EBL. Evolving galaxy populations compose the bulk of the EBL which is dominated by starlight in the UV/optical and thermally radiating dust at longer IR wavelengths. We use multiwavelength survey data to fit the evolution of a single derived quantity, the luminosity density, in this otherwise assumption-free approach. We cover the whole range from UV to mid-IR (0.15-25\mic), providing for the first time an empirical derivation of the $\gamma\gamma$ optical depth out to several TeV. \citet{Stecker12} have used a similar approach to reconstruct the EBL at $<0.7$\mic. The survey data used in this Letter extends the collection of LFs presented in \citet{Helgason12} with expanded coverage in the UV \citep{Steidel99,Sullivan00,Treyer05,Schiminovich05,Budavari05,Sawicki06,Tresse07,Finkelstein12} and out to mid-IR wavelengths \citep{Rujopakarn10,Shupe98,Xu98,Sanders03,LeFloch05,Magnelli11,PerezGonzalez05}. This library now contains \NLF\ measured LFs and allows us to reconstruct the evolving EBL and its spectrum in a finely sampled wavelength grid encompassing 0.15-24\mic\ out to $z\sim 2-8$. \citet{Helgason12} used this data to accurately recover the observed galaxy number counts in the 0.45-4.5\mic\ range. This therefore assumes only the existence of populations now observed out to $z\sim 2-8$ and magnitudes as faint as $m_{\rm AB}\sim 23-26$; additional populations, such as the hypothetical galaxies with first stars would then exist at still earlier times and have much fainter fluxes.
\begin{figure}[t]
  \begin{center}
      \includegraphics[width=0.45\textwidth]{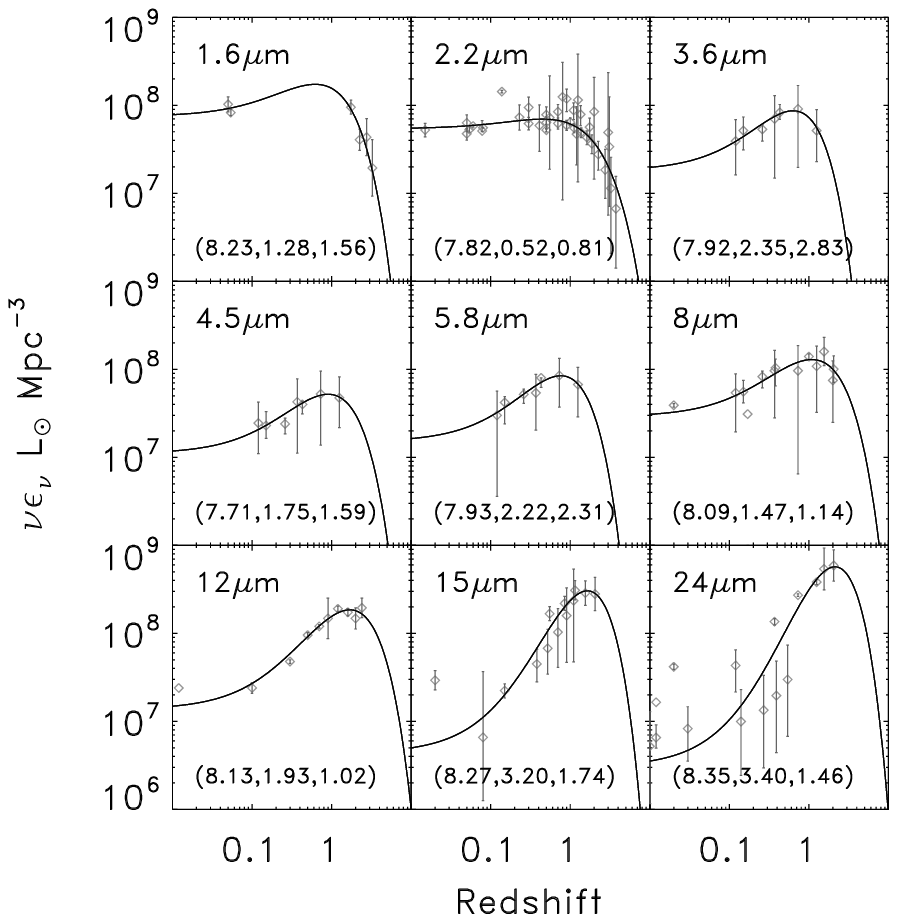}
  \end{center}
      \caption{ Same as Figure~\ref{fig_ld1} for infrared wavelengths, 1.6-24\mic. The references for the data points can be found in Table 1 of \citet{Helgason12} with the addition of mid-IR data from \citet{Shupe98,Xu98,Sanders03,LeFloch05,PerezGonzalez05,Rujopakarn10,Magnelli11}. }
      \label{fig_ld2}
\end{figure}

In the rest-frame UV to near-IR, the LF is well described by the conventional Schechter parameterization whereas at mid-IR wavelengths, the LF seems to be better described by a broken power-law or a double-exponential profile. Regardless of the functional form, the LFs can be integrated to give the comoving volume emissivity (we refer to this as the luminosity density) in the given rest-frame band
\begin{equation} 
   \epsilon_\nu(z) = \int L_\nu\phi(L_\nu,z)dL_\nu
\end{equation}
where $\phi(L_\nu,z)$ is the measured LF. Each data point in Figures~\ref{fig_ld1} and \ref{fig_ld2} represents the luminosity density given by the authors along with 1$\sigma$ error bars (for references, see Table 1 in \citet{Helgason12}). In the cases where this value is not given explicitly in the original papers, we have integrated the best-fit parameterized LF to obtain $\epsilon_\nu$ and have estimated the errors from the distribution of all of the values of $\epsilon_\nu$ allowed within the 1$\sigma$ solutions of the individual fit parameters. However, mutual comparison of uncertainties among the many different studies is not very meaningful since some authors include various effects in addition to the statistical errors from the method of LF estimation, such as cosmic variance,  $k$-corrections, incompleteness and photometric system. Here, we have chosen to maximize our wavelength and redshift coverage by letting all available measurements contribute to our fitted evolution regardless of the error treatment. In the cases where the median redshift of the sample is not explicitly given, we have placed the measurements at the midpoint of the redshift bin. Our wavelength interval is sampled at the rest-frame bands shown in the panels of Figure~\ref{fig_ld1} and \ref{fig_ld2} where most of the LFs have been measured. The offsets from these defined wavelengths due to filter variations (e.g. SDSS $u^\prime$ and Johnson $U$) are small enough to be neglected.
\begin{figure*}
  \begin{center}
      \includegraphics[width=0.53\textwidth]{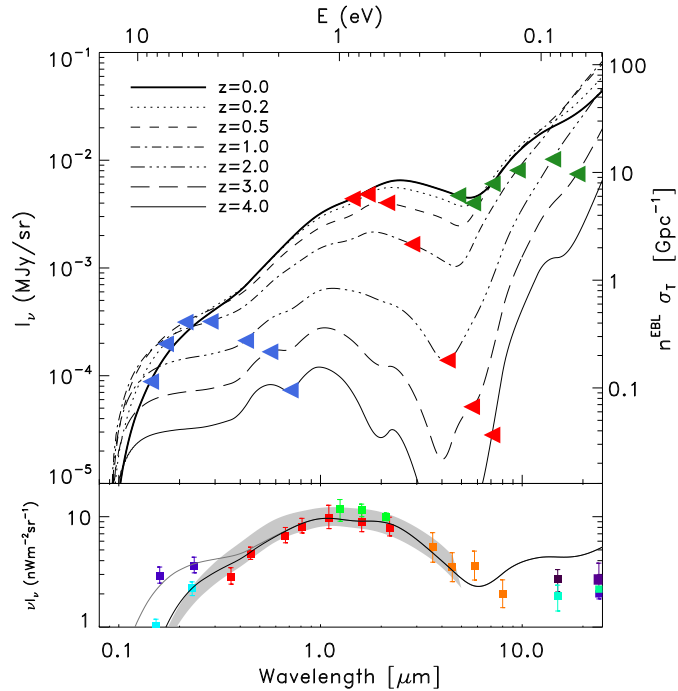}
  \end{center}
      \caption{ Upper: the evolving EBL resulting from our empirical reconstruction. The different lines illustrate the build-up of EBL with cosmic time leading to the present day levels (thick solid curve). We display the evolving EBL in comoving coordinates i.e. without the expansion factor $(1+z)^3$ for better mutual comparison. The left pointing triangles show the EBL threshold energy needed to interact with an observed $30 {\rm GeV}$ photon (blue), $300 {\rm GeV}$, (red) and $1 {\rm TeV}$ (green) originating at the redshifts shown. Lower: our reconstructed EBL compared to integrated counts in the literature along with the region bound by the upper/lower faint-end scenarios in \citet{Helgason12} (shaded). The black and gray lines represent the case of $E_{\rm cutoff}=10.2{\rm eV}$ and $13.6{\rm eV}$ respectively. The counts data are from \citep{Gardner00} (purple), \citep{Xu05} (cyan), \citep{Madau&Pozzetti00} (red), \citep{Keenan10a} (green), \citep{Fazio04} (orange), \citep{Hopwood10} (cyan), \citep{Metcalfe03} (black), \citep{Papovich04} (deep purple), \citep{Bethermin10} (light green), \citep{Chary04} (blue).  }
      \label{fig_nebl}
\end{figure*}

Motivated by the fitting formulae in \citet{Helgason12}, which we found to be reliable over a wide range of wavelengths, we consider the following three parameter fits for the evolution of the luminosity density
\begin{equation} \label{eqn:ldfit}
   \epsilon_\nu(z) = a_\lambda (1+(z-z_0))^{b_\lambda}\exp{\left( -c_\lambda(z-z_0) \right)}
\end{equation}
where we fix $z_0=0.8$. Although not restricted to Schechter LFs, this functional form for $\epsilon_\nu(z)$ is nevertheless equivalent to the underlying Schechter parameters evolving as $L^\star \propto (1+(z-z_0))^{b_\lambda}$ and $\phi^\star \propto \exp{\left(-c_\lambda(z-z_0)\right)}$ with a constant faint-end slope. Interpolating the rest-frame data between our 18 reference bands defines the rest-frame emissivity spectrum at any given epoch. We apply a cutoff to the spectrum above the Lyman limit, $E_{\rm cutoff} = 13.6 {\rm eV}$, corresponding to efficient absorption of ionizing photons by hydrogen in the local environments. At all lower energies, the universe is assumed to be completely transparent to background photons. The integrated light from galaxies seen today is (e.g. Peebles 1971)
\begin{equation}
  \nu I_\nu = \frac{c}{4\pi} \int_0^\infty  \nu^\prime \epsilon_{\nu^\prime}(z)  \left|\frac{dt}{dz}\right|\frac{dz}{(1+z)},
\end{equation}
where $\nu^\prime = \nu (1+z)$ is the rest-frame frequency and
\begin{equation}
  \left|\frac{dt}{dz}\right| = \frac{1}{H_0(1+z)\sqrt{(1+z)^3\Omega_m + \Omega_\Lambda }}.
\end{equation}
In \citet{Helgason12} we considered two limiting cases for the evolution of the faint-end of the LF and showed that the distribution of galaxies from LF data, when projected onto the sky, accurately recovered the observed galaxy counts across the optical and near-infrared. The flux from the integrated counts is displayed as shaded regions in the lower panel of Figure~\ref{fig_nebl} along with our empirically determined EBL (solid line) which is in good agreement with integrated counts data in the literature apart from wavelengths $\gsim 6{\rm \mu m}$ where the steep evolution of the mid-IR LFs (8-24\mic) causes our EBL to be a factor of $\sim$2-3 higher than integrated counts from \citet{Chary04}, \citet{Hopwood10}, \citet{Papovich04} and \citet{Bethermin10}. Although these authors do not claim to fully resolve the CIB at these wavelengths, the discrepancy is large enough to indicate a mismatch between number counts and mid-IR LF measurements at $z > 0.5$. This issue is apparently also encountered in galaxy evolution models; \citet{Somerville12} are not able to simultaneously account for the integrated counts and the bright-end of the observed LF in the 8-24\mic\ range using different dust templates. In fact, this is the wavelength regime where varying degrees of dust contribution and PAH emission make the spectrum less predictable. Recent upper limits derived from a TeV source spectra also favor low levels of CIB at these wavelengths \citep{Orr11,Meyer12}. At this stage, one must therefore question the robustness of EBL reconstructed from LFs at $\gsim 8{\rm \mu m}$.

The UV/blue end of the EBL turns out to be sensitive to the abundance of photons with energies just below the Lyman limit, 13.6eV. The redshifted far-UV contribution dominates the EBL below 0.5\mic\ due to the steep increase of the star formation rate at earlier times. For all galaxy types, there is considerable absorption in the Lyman series which we do not account for and we illustrate this dependence by considering the case where Lyman-series absorption completely suppresses the spectrum above $10.2 {\rm eV}$ (instead of $E_{\rm cutoff}=13.6 {\rm eV}$); shown as gray lines in Figure~\ref{fig_nebl} (lower panel). We subsequently display our optical depths for the both cases which bracket the true behavior.

\section{The Photon-photon Optical Depth}

\begin{figure}
  \begin{center}
      \includegraphics[width=0.45\textwidth]{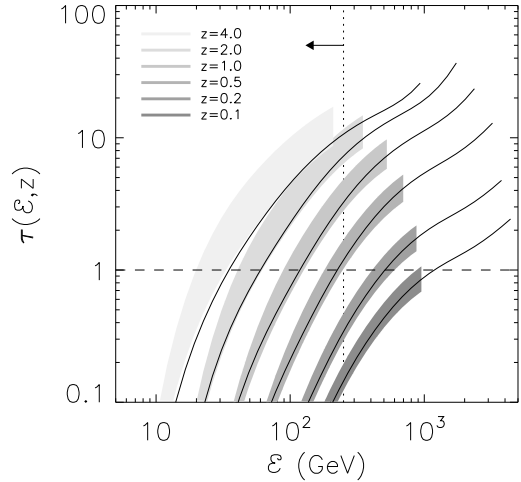}
  \end{center}
      \caption{ Solid lines show the $\gamma\gamma$ optical depth contributed by known galaxy populations assuming $E_{\rm cutoff}=13.6{\rm eV}$ (gray curve in the lower panel of Figure~\ref{fig_nebl}). The curves are not drawn beyond the energy of $(m_ec^2)^2/E_{\rm 24\mu m}(1+z)$ as we do not consider data at $\lambda>25$\mic. The shaded regions show the boundaries of the upper/lower scenarios in the empirical reconstruction of \citet{Helgason12} out to 4.5\mic. The dotted vertical line shows roughly the highest energy probed by {\it Fermi}/LAT and the dashed line shows where $\tau=1$ for reference. }
      \label{fig_tau}
\end{figure}

The relevant quantity for computing the optical depth due to photon-photon interaction is the rest-frame number density of photons as a function of time and energy, $n(E,z)$. We shall refer to the energy of a photon belonging to the EBL as $E$ and we use $\mathcal{E}$ for the propagating $\gamma$-rays. Rest-frame quantities are denoted with a prime. At any given epoch, the photon number density (in proper coordinates) is composed both of sources emitting in the rest-frame as well as the accumulated emission from earlier times
\begin{equation} \label{eq_n}
  n(E^\prime,z) 
                = (1+z)^3 \int_{z}^\infty  \frac{\epsilon_{\nu^\prime}(z^\prime)/h}{h\nu^\prime} \frac{dt}{dz^\prime}dz^\prime,
\end{equation}
where $h$ is the Planck constant (the extra $h$ is to convert $\epsilon_\nu$ to per unit energy, $dE=hd\nu$) and $\nu^\prime=(1+z^\prime)/(1+z)$. The condition for pair production is that the total energy in the center-of-mass frame must satisfy $\mathcal{E^\prime}E^\prime(1-\cos\theta) \geq 2(m_ec^2)^2$ where $\theta$ is the angle of incidence.
This means that in order to interact with a $\gamma$-ray of energy $\mathcal{E}^\prime$, background photons must have wavelengths of $\mathrm{\leq 1.0 (\mathcal{E}^\prime/210{\rm GeV}){\rm \mu m}}$. The cross section for this interaction is
\begin{equation}
  \sigma(E^\prime,\mathcal{E}^\prime,\mu) = \frac{3\sigma_T}{16}(1-\beta^2)\left[2\beta(\beta^2-2) + (3-\beta^4) \ln{\left( \frac{1+\beta}{1-\beta}\right)} \right],
\end{equation}
where
\begin{equation*}
  \beta = \sqrt{ 1 - \frac{2m_e^2c^4}{ E^\prime\mathcal{E}^\prime(1-\mu) } }, \hspace{15pt} \mu = \cos{\theta}.
\end{equation*}
For the most likely angle of incidence, $\mu=0$ (side-on), the probability for interaction is maximized at roughly four times the minimum threshold energy, $\sim 4m_ec^2/\mathcal{E}$. The optical depth encountered by a high energy photon originating at $z$ ($\tau \! \sim \! \sigma nl$) can be expressed in terms of its observed energy, $\mathcal{E}$, as
\begin{equation} \label{eq_tau}
  \begin{split}
    \tau_{\gamma \gamma}(\mathcal{E},z) = c\int_0^{z} & \frac{dt}{dz^\prime}dz^\prime  \int_{-1} ^1(1-\mu)\frac{d\mu}{2} \\
     & \times \int^\infty_{ 2m_e^2c^4/\mathcal{E}^\prime(1-\mu) }  \sigma(E^\prime,\mathcal{E}^\prime,\mu) n(E^\prime,z^\prime) dE^\prime
   \end{split}
\end{equation}
where $n(E,z)$ comes from Equation \ref{eq_n}. In Figure~\ref{fig_tau} we display the calculated optical depths as a function of observed energy for $\gamma$-rays originating at several redshifts. The optical depth roughly traces the shape of the number density of EBL photons with a sharp drop in the optical depth at the lowest energies due to the cutoff at the Lyman limit. We also show the regions encompassed by the two scenarios of the faint-end evolution from the reconstruction of \citet{Helgason12}.

\section{ Application to High Energy Observations }

\begin{figure}
  \begin{center}
      \includegraphics[width=0.45\textwidth]{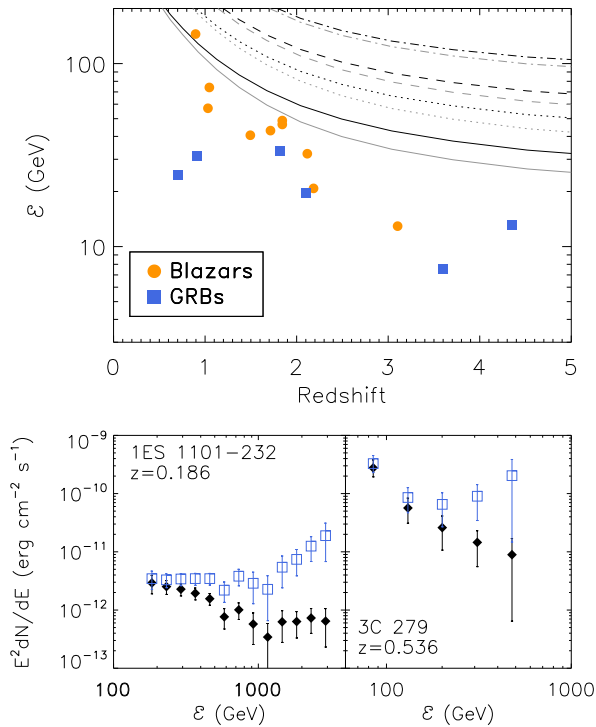}
  \end{center}
      \caption{ Upper: the curves show where a photon of energy $\mathcal{E}$ originating at $z$, encounters exactly $\tau_{\gamma\gamma}=1,2,3,5$ (solid, dotted, dashed, dash-dotted respectively). Black and gray correspond to the cases of the rest-frame spectrum cutoff energy is at 10.2 and 13.6eV respectively. The symbols show the highest energy photon observed in GRBs (blue squares) and a selection of the most constraining of {\it Fermi}/LAT blazars (orange circles) \citep{Abdo10,McConville11}. Lower: the observed energy spectrum of the sources 1ES 1101-232 ($z=0.186$; {\it left}) and 3C 279 ($z=0.536$; {\it right}) shown with black diamonds, and the corresponding deabsorbed data as blue squares. The deabsorbed spectra have best-fit photon indices $\Gamma_{\rm in}=1.49$ and $\Gamma_{\rm in}=2.28$ respectively, but deviate substantially from a power-law at the highest energy bins.  }
      \label{fig_fermi}
\end{figure}

What do these reconstructed opacities imply for observations of high-energy sources with the current instruments? Blazars and gamma-ray bursts (GRBs) are examples of high energy extragalactic sources whose spectra is affected by attenuation of photons in excess of $\gsim 10 {\rm GeV}$. Two types of datasets are relevant for this discussion: space-borne {\it Fermi}/LAT measurements at $\lsim 300$GeV and ground-based telescopic measurements extending out to $\sim$TeV energies. Because extragalactic $\gamma$-ray absorption increases with both with redshift and energy, the EBL can be constrained based on the highest energy photons observed from a source provided the redshift is known \citep{Abdo10}. GRBs have the advantages of being observable across great distances and typically displaying harder spectra than most blazars at sub-GeV energies. Figure~\ref{fig_nebl} shows that for {\it Fermi}-observed sources it is sufficient to use data out to $\simeq4.5$\mic\ (red symbols), whereas for TeV range observations, survey data are needed out to the longer wavelengths (green symbols). 
Figure \ref{fig_tau} shows the reconstructed optical depth explicitly confirming this. 
We now briefly discuss the implications:

{\bf Fermi/LAT} detects blazars and GRBs out to energies $\sim 250-300$ Gev. In Figure~\ref{fig_fermi} (upper panel) we show curves of constant $\gamma\gamma$ optical depth in $\mathcal{E}-z$ space and compare with the most constraining high energy {\it Fermi}/LAT sources with known redshifts \citep[taken from][]{Abdo10}. The contours of $\tau=1,2,3,5$ correspond to probabilities of photon being absorbed by the EBL of 63\%, 86\%, 95\%, 99.3\%, respectively. In the absence of new populations, the universe remains fairly transparent at the {\it Fermi}/LAT energies out to $z\sim 2-3$. Our reconstructed EBL is fully consistent with all the available LAT data and, in fact, allows for non-negligible extra levels of the CIB from new populations such as possibly have existed at higher $z$. As the Fermi mission progresses and Figure~\ref{fig_fermi} (upper) becomes more populated at the highest energies, sources at high-$z$ will provide better constraints for the optical/NIR EBL.

{\bf Ground-based Cherenkov telescopes} have produced good quality spectra for TeV-blazars, although for sources at significantly lower redshifts than {\it Fermi}/LAT.
If the evolving EBL is known to a good accuracy, one can deabsorb observed blazar spectra to reveal the intrinsic component, which is expected to have a power-law form, $dN/dE\propto E^{-\Gamma}$. The lower panels in Figure~\ref{fig_fermi} demonstrate how our reconstructed EBL affects the spectrum of two known blazars at relatively high redshifts, both of which have been used to place upper limits on the optical/NIR EBL. Good quality spectrum of the BL Lac object 1ES 1101-232 ($z=0.186$) has been obtained by HESS in the energy range $0.16-3.3 {\rm TeV}$ which at $z=0.186$ interacts most strongly with optical and near-IR background photons \citep{Aharonian06}. The observed spectrum is relatively hard ($\Gamma=2.88\pm 0.17$) and results in a best-fit intrinsic photon index of $\Gamma_{\rm int}=1.49$ after deabsorption. For this particular source, the upturn at TeV energies is largely driven by the EBL photons at $\gsim 5$\mic\ and would be less pronounced if our EBL reflected the integrated counts data in the lower panel of Figure~\ref{fig_nebl}.
Because the EBL changes with time in both shape and amplitude, the effects on $\gamma$-ray absorption become even more prominent for more distant sources. The spectrum of the distant radio quasar 3C 279 has been captured by MAGIC during different flaring events \citep{Albert08,Aleksic11}. The deabsorbed spectrum of 3C 279 shown in Figure~\ref{fig_fermi} (lower) also deviates substantially from a simple power-law in the highest energy bins which is unlikely to be due to our near-IR background being overestimated. This behavior of the deabsorbed 3C 279 spectrum has been pointed out by \citet{Dominguez11} who suggest either improved emission models or instrumental systematic uncertainties as potential solutions. Another possibility is that some fundamental effects are missing such as secondary $\gamma$-rays produced along the line of sight by cosmic rays accelerated by the blazar jet \citep{EsseyKusenko10}. In absence of such secondary effects however, the universe should be completely opaque for TeV sources at $z \gsim 0.5$.

Upper limits for the EBL derived from TeV spectra rely on assumptions of the hardness of the intrinsic blazar spectrum. \citet{Meyer12} derive limits for the whole range of optical to far-IR EBL using an extensive source sample from both {\it Fermi}/LAT and ground-based Cherenkov telescopes. Their results allow a total integrated NIR flux (1-10\mic) of $\sim$20\nW\ in excess of known galaxy populations whereas there is, at most, little room for extra contribution in the mid-IR ($>10$\mic). In fact, our LF-derived EBL is inconsistent with the lowest mid-IR limits of \citet{Orr11}.

\section{Summary}

We have shown that it is possible to robustly reconstruct the evolving EBL in the universe using our earlier library of multiwavelength survey data now updated to extend from the UV out to the mid-IR \citep{Helgason12}. This reconstruction uniquely defines the $\gamma$-ray opacity out to TeV energies for sources at $z\lsim 4$ and shows that at the energy bands probed by {\it Fermi}/LAT, the universe is fairly transparent out to $z\sim 2-3$, unless unknown sources at high redshifts contribute non-negligible amounts of CIB. At TeV energies, probed by ground based telescopes, the universe becomes optically thick at $z\sim 0.5$ so any such photons associated with the sources at higher redshifts would have to be of secondary origin. 

Our reconstructed EBL and optical depths are available upon request. 
This work was supported by NASA Headquarters under the NASA Earth and Space Sciences Fellowship Program - Grant NNX11AO05H. KH is also grateful to {\it The Leifur Eiriksson Foundation} for its support. We thank W. McConville, B. Magnelli and M. Ricotti for useful communications.


\label{lastpage}
\end{document}